\documentclass[11pt]{article}

\usepackage{amsmath, amssymb, bm}
\usepackage{graphicx}
\usepackage{geometry}
\usepackage{hyperref}
\usepackage{natbib} 
\title{\bf Polyphonic Intelligence:\\
Constraint-Based Emergence, Pluralistic Inference, and Non-Dominating Integration}

\author{
Alexander D. Shaw\\
{\small Computational Psychiatry \& Neuropharmacological Systems (CPNS) Lab}\\
{\small Department of Psychology, Faculty of Health \& Life Sciences}\\
{\small University of Exeter, UK}\\
{\small \href{https://cpnslab.com}{https://cpnslab.com}}\\
{\textit{\small \href{a.d.shaw@exeter.ac.uk}{a.d.shaw@exeter.ac.uk}}}
}

\date{}

\begin{document}
\maketitle

\begin{abstract}
Across neuroscience, artificial intelligence, and related fields, dominant models of intelligence typically privilege convergence: uncertainty is reduced, competing explanations are eliminated, and behaviour is governed by the optimisation of a single objective or policy. While this framing has proved powerful in many settings, it sits uneasily with biological and adaptive systems that maintain redundancy, ambiguity, and parallel explanatory processes over extended timescales.

Here we propose an alternative perspective, termed polyphonic intelligence, in which coherent behaviour and meaning emerge from the coordination of multiple semi-independent inferential processes operating under shared constraints. Rather than resolving plurality through dominance or collapse, polyphonic systems sustain multiple explanatory trajectories and integrate them through soft alignment, compatibility relations, and bounded influence.

We develop this perspective conceptually and formally, introducing a variational framework in which multiple coordinated approximations are maintained without winner-takes-all selection. This formulation makes explicit how plurality can remain stable, tractable, and productive, and clarifies how polyphonic inference differs from ensemble methods, mixture models, and Bayesian model averaging.

Through proof-of-principle examples, we demonstrate that non-dominating, pluralistic inference can be implemented in simple computational systems without requiring centralised control or global convergence. We conclude by discussing implications for neuroscience, psychiatry, and artificial intelligence, and by arguing that intelligence may be more fruitfully understood as coordination without command rather than as the elimination of uncertainty.
\end{abstract}

\section{Introduction}

Despite substantial differences in surface-level formalisms across neuroscience, artificial intelligence, cognitive science, control theory, and statistical physics, remarkably similar assumptions about intelligence recur across these domains. Intelligence is most often characterised in terms of reducing uncertainty, minimising error, optimising performance, or converging on a solution, such that behaviour is formalised as optimisation toward a single explanatory state, representation, or policy. Whether articulated through Bayesian inference, optimal control, reinforcement learning, or efficient coding, this orientation toward convergence has become a common organising principle \citep{mackay2003information,bishop2006pattern,friston2010free,lake2017building,dayan2001theoretical,todorov2002optimal}.

There are clear reasons why such a framing has proven influential. Much of modern machine learning, signal processing, and computational neuroscience is built upon it, and many theoretical and practical advances would not have been possible without the simplifying assumptions it affords \citep{bishop2006pattern,mackay2003information,lake2017building,russell2021artificial,wolpert1997no}. At the same time, the success of these approaches has normalised a normative assumption that is rarely stated explicitly, namely that intelligence is fundamentally convergent and that uncertainty, multiplicity, and ambiguity are primarily problems to be resolved. Whether this assumption continues to hold as models move beyond idealised or engineered settings, particularly when biological and adaptive systems are considered, remains an open question.

When viewed through the lens of biological intelligence, strict convergence appears more as an exception than as a rule. Neural systems routinely maintain redundant and overlapping representations, multiple perceptual interpretations can coexist across cortical hierarchies, and action selection often unfolds in the presence of unresolved alternatives \citep{sporns2011networks,tononi1994measure,breakspear2017dynamic,fusi2016mixed,decharms2000neural}. Rather than collapsing uncertainty in a single step, learning processes tend to reshape representational landscapes gradually and often incompletely \citep{clark2013whatever,friston2010free}. Across levels of organisation, cognition therefore appears able not only to tolerate persistent multiplicity, but in some cases to exploit it \citep{kelso1995dynamic,breakspear2017dynamic}.

Taken together, these observations invite a reconsideration of convergence as a defining property of intelligence. Instead of viewing intelligent behaviour primarily as the optimisation of a single global objective, it may be more naturally understood as a process of coordination among multiple interacting processes \citep{beer2000dynamical,kelso1995dynamic,varela1991embodied}. We refer to this alternative framing as \emph{polyphonic intelligence}.

The term “polyphonic” is used deliberately to emphasise the coexistence of multiple semi-independent processes whose interactions give rise to coherent structure without requiring their suppression. This usage is intended in a precise sense rather than as a loose metaphor. As we show throughout this paper, closely related mathematical and dynamical principles recur across inference, control, neural systems, and physical models of collective behaviour \citep{strogatz2003sync,breakspear2017dynamic,olfati2007consensus,couzin2005effective}. Before developing this perspective further, we first situate it within the historical and theoretical developments that have shaped contemporary models of intelligence.

\section{Historical and Theoretical Context}

\subsection{Convergence as an implicit design principle}

Despite substantial differences in surface-level formalisms across neuroscience, artificial intelligence, control theory, and related fields, a shared orientation toward convergence has gradually taken hold, often without being articulated as a core theoretical commitment. In Bayesian inference, for example, uncertainty is represented probabilistically, yet practical inference is frequently reduced to point estimates such as maximum a posteriori solutions or posterior means, particularly when tractability or computational efficiency is prioritised \citep{mackay2003information,bishop2006pattern,gelman2013bda,blei2017variational,wainwright2008graphical}. A closely related tendency emerges in variational inference, where intractable posteriors are replaced by simpler approximating families that favour unimodal, factorised, or otherwise collapsed representations, with the effect of privileging a single explanatory trajectory.

Similar assumptions can be traced through control-theoretic and reinforcement learning frameworks, albeit expressed through different mathematical idioms. Within classical optimal control, intelligence is formalised in terms of policies that minimise expected cumulative cost, while reinforcement learning frames behaviour through value functions whose maximisation defines optimal action selection \citep{Sutton1998,bertsekas2017dp,todorov2002optimal}. Although these approaches differ in their origins and technical machinery, multiplicity is typically treated as a transient feature of exploration or uncertainty, expected to resolve as learning proceeds and a stable solution is reached.

That such convergent formulations have become deeply embedded in formal models of intelligence is therefore unsurprising. Optimisation problems are often analytically tractable, admit convergence guarantees under well-defined conditions, and lend themselves readily to engineering implementation; over time, these advantages have allowed convergence to be treated less as a modelling choice and more as a defining property of intelligent systemsn \citep{wolpert1997no}.

\subsection{Why convergent formulations have been successful}

In many of the domains that shaped early theories of intelligence, the assumption that uncertainty could be reduced to a single best explanation was not only defensible but highly productive. Signal processing, system identification, supervised learning, and classical perception tasks frequently involve environments that are stationary, low-dimensional, or well approximated by convex objectives, conditions under which convergent inference performs reliably and efficiently \citep{boyd2004convex,kalman1960new,hastie2009elements}.

Within such settings, collapsing ambiguity into a single actionable estimate simplifies decision-making, stabilises control, and supports efficient learning. Convergent solutions are also comparatively easy to interpret, communicate, and deploy, contributing to their appeal across scientific and engineering contexts. The prominence of optimisation-based approaches in machine learning and computational neuroscience therefore reflects substantial empirical and practical success rather than theoretical conservatism \citep{boyd2004convex,hastie2009elements}.

As these methods have been extended to increasingly complex problem domains, however, the same assumptions have often been carried forward with little modification. When systems become high-dimensional, non-stationary, or tightly coupled to their environments, the expectation that intelligent behaviour necessarily entails convergence becomes progressively harder to sustain.

\subsection{Limits of convergence in complex systems}

When attention shifts toward biological and adaptive systems, a number of tensions with strict convergence become difficult to ignore. Many inference problems in such settings are fundamentally ill posed, admitting multiple explanations that are equally consistent with available data, such that multimodal posteriors, parameter degeneracy, and non-identifiability arise as generic features rather than exceptional cases \citep{tarantola2005inverse,gutenkunst2007universally,raue2009structural}.

Redundancy and overlap are pervasive in biological organisation. Neural populations often encode similar information in different ways, while learning tends to reshape representational landscapes gradually rather than collapsing them decisively. Behaviour, moreover, unfolds in environments that are non-stationary, partially observable, and influenced by the agent’s own actions \citep{kaelbling1998planning}, further undermining the assumption that a single optimal solution exists or can be maintained over extended timescales.

Under these conditions, enforcing convergence can lead to brittle behaviour, premature commitment, or a loss of adaptive flexibility. Maintaining multiple, partially incompatible hypotheses may instead support resilience and rapid reconfiguration, allowing intelligent systems to respond effectively as circumstances change. From this perspective, convergence appears less as a universal hallmark of intelligence and more as one strategy among several.

\subsection{Existing responses to non-convergence}

In response to these challenges, a variety of approaches have attempted to incorporate plurality within formal models of inference and control. Mixture models and ensemble methods retain multiple hypotheses in parallel, while Bayesian model averaging integrates over competing models rather than selecting a single best one. Within neuroscience, population coding and distributed representations explicitly avoid single-unit dominance, and dynamical systems theory has emphasised metastability and transient coordination in place of fixed-point convergence \citep{kelso1995dynamic,strogatz2003sync,dietterich2000ensemble,hoeting1999bma,mclachlan2000finite,rabinovich2008neuro}.

Even within these frameworks, however, plurality is often treated as provisional rather than stable. Ensembles are typically averaged, mixture components are weighted toward dominance, and metastable dynamics are frequently analysed in terms of transitions between attractors. As a result, inference and decision-making continue to resolve toward a single outcome at the level of action, explanation, or control \citep{hinton2015distilling}.

What remains largely unexplored is the possibility that plurality itself may constitute a stable and productive mode of organisation rather than a temporary stage en route to convergence. While existing frameworks relax convergence at the level of representation, they tend to reinstate it at the level of integration, leaving open the question of how coordination might be achieved without dominance.

\section{Polyphonic Intelligence: Core Concepts}

Having argued that many dominant models of intelligence implicitly privilege convergence, and that existing attempts to accommodate multiplicity often reintroduce dominance at the level of integration, we now turn to the core conceptual elements of an alternative framing, which we refer to as \emph{polyphonic intelligence}. Within this framework, plurality is not treated as a transient inconvenience to be resolved, but as a stable and potentially productive mode of organisation. Rather than emerging from the optimisation of a single global objective, intelligent behaviour is understood as arising through the coordination of multiple semi-independent processes operating under shared constraints.

\subsection{Voices as semi-independent inferential processes}

Central to this framework is the notion of a \emph{voice}. By a voice, we mean a semi-independent inferential or dynamical process that maintains its own internal state and generates predictions or hypotheses about the world. Voices need not share the same structure or scope. They may differ in their priors, parameterisations, temporal horizons, or representational commitments, reflecting alternative ways of interpreting or engaging with the same environment. What distinguishes a voice is, therefore, not its optimality with respect to a single criterion, but its coherence as an internally consistent process capable of responding to evidence.

Importantly, voices are not intended to function as mere samples from a single posterior distribution, nor as redundant replicas of an underlying model. Instead, each voice traces a distinct explanatory trajectory through model space, potentially emphasising different aspects of the data or environment. In biological systems, such plurality is reflected in overlapping neural representations, parallel processing streams, and competing perceptual interpretations that coexist over time \citep{rao1999predictive,friston2017active,rigotti2013mixed,spivey2007continuity}. In artificial systems, voices may correspond to alternative generative models, parameter regimes, policies, or hypotheses that are maintained in parallel rather than being immediately resolved.

From this perspective, voices are more naturally understood as \emph{perspectives} than as candidates in a competition. Agreement is not required, nor are voices ranked solely according to a scalar notion of fit. Instead, they coexist for as long as they remain viable under the constraints imposed by the system and its environment \citep{beer2000dynamical,friston2017active}.

\subsection{Constraints, compatibility, and coordination}

Although polyphonic systems involve multiple semi-independent processes, they are not arbitrary collections of disconnected components. Coherence arises because voices are embedded within a shared constraint structure. Such constraints may be imposed by sensory data, environmental feedback, physical laws, task demands, or internal resource limitations, shaping the space of possible interactions among voices without enforcing agreement directly \citep{haken1983synergetics,olfati2007consensus,kelso1995dynamic}.

Within this setting, coordination does not depend on centralised control or explicit arbitration; instead, it emerges through compatibility. Voices that generate mutually consistent predictions, actions, or interpretations tend to align more closely, while those that are incompatible may remain weakly coupled or transiently decouple. Coordination is therefore inherently soft, allowing degrees of alignment rather than binary acceptance or rejection \citep{kelso1995dynamic,barandiaran2009defining}.

Crucially, constraints operate at the level of relations between voices rather than solely at the level of individual accuracy. A voice may persist even when it is locally suboptimal, provided it contributes to a coherent configuration that satisfies global constraints. This stands in contrast to convergent frameworks, in which subdominant hypotheses are rapidly eliminated once a better-fitting alternative is identified.

\subsection{Non-dominating integration}

Within polyphonic intelligence, coordination among voices is achieved through what we term \emph{non-dominating integration}. Rather than collapsing plurality into a single dominant explanation, multiple voices continue to influence inference, prediction, and action \citep{tognoli2014metastable,cisek2007cortical} without any single process exerting absolute control.

Non-dominating integration does not imply equal weighting or indecision. Voices may exert different degrees of influence, and their relative contributions may change over time. However, influence is bounded in such a way that no individual voice is permitted to fully suppress the others. This prevents premature commitment, preserves exploratory capacity, and allows the system to remain responsive to novel or conflicting evidence \citep{bogacz2006physics}.

This principle distinguishes polyphonic intelligence from ensemble averaging, mixture models, and model selection approaches because in those frameworks, plurality is typically resolved through weighting schemes that converge toward dominance. By contrast, non-dominating integration treats plurality as a persistent feature of intelligent organisation, with coordination achieved through soft alignment rather than elimination.

\subsection{Viability versus optimality}

Underlying these concepts is a shift in the criteria by which intelligent behaviour is evaluated. Convergent frameworks typically assess success in terms of optimality, defined as the minimisation of error, cost, or free energy with respect to a single objective. Polyphonic intelligence instead places emphasis on \emph{viability} \citep{aubin1991viability,ashby1956design}. A configuration of voices is considered viable if it remains internally coherent, responsive to constraints, and capable of adaptive coordination over time.

Optimisation is not rejected outright within this view, but its role is relativised. Local optimisation may occur within individual voices, while global behaviour is governed by the maintenance of a viable plural configuration rather than by convergence to a unique solution \citep{dipaolo2010enactive}. This distinction becomes particularly salient in non-stationary, ambiguous, or open-ended environments, where rigid optimisation can undermine adaptability.

By reframing intelligence as coordination-without-command, polyphonic intelligence provides a conceptual foundation for systems that tolerate ambiguity, sustain multiple explanations, and adapt through soft constraint satisfaction rather than hard convergence. In the following section, these ideas are formalised within a variational framework, making explicit how voices, constraints, and non-dominating integration can be implemented computationally.

\section{Formal Framework: Polyphonic Variational Inference}

This section describes a formal instantiation of polyphonic intelligence within a variational inference setting. The aim here is not to enumerate all possible implementations, but to make explicit a minimal mathematical structure in which plural, non-dominating inference can be defined precisely, analysed, and distinguished from existing convergent formulations. By situating polyphony within a familiar variational framework, the goal is to retain tractability while altering the assumptions that govern integration and collapse.

\subsection{Single-model variational inference and the logic of collapse}

To provide a point of reference, it is useful to briefly recall the standard variational inference framework. Let $y$ denote observed data and $x$ latent variables governed by a generative model $p(y,x)=p(y\mid x)p(x)$. Variational inference proceeds by approximating the true posterior $p(x\mid y)$ with a tractable distribution $q(x)$, chosen to minimise the variational free energy
\begin{equation}
F[q] = \mathbb{E}_{q(x)}[\log q(x) - \log p(y,x)],
\end{equation}
which provides an upper bound on the negative log model evidence \citep{neal1998view,jordan1999variational,wainwright2008graphical,blei2017variational}.

In practice, the approximating distribution $q(x)$ is drawn from a restricted family, such as Gaussian or mean-field forms, and its parameters are adjusted to minimise $F$. Although the underlying formulation is probabilistic, the resulting optimisation dynamics tend to favour convergence toward a single mode of the posterior, particularly when unimodal approximations or point-estimate summaries are employed. As optimisation proceeds, uncertainty is therefore progressively reduced and competing explanations are eliminated \citep{turner2011two,blei2017variational,wainwright2008graphical}.

Rather than reflecting a deficiency of variational inference itself, this behaviour follows directly from the structure of the objective being optimised. A single functional is minimised with respect to a single approximating distribution, leaving little room for persistent plurality. Polyphonic inference departs from this structure by replacing the single approximation with a coordinated family of approximations.

\subsection{Polyphonic generative models and voices}

Within a polyphonic setting, inference is no longer carried out by a single variational distribution, but by a collection of $K$ distributions $\{q_k(x)\}_{k=1}^K$, each corresponding to a distinct voice. All voices are conditioned on the same observations $y$, yet they may differ in their priors, parameterisations, initial conditions, or internal dynamics, reflecting alternative explanatory trajectories through model space \citep{jordan1999variational,gershman2012nonparametric}.

Each voice (k) is associated with its own variational free energy,
\begin{equation}
F_k = \mathbb{E}_{q_k(x)}[\log q_k(x) - \log p(y,x)],
\end{equation}
which quantifies the internal consistency of that voice with respect to the generative model and the data. Crucially, this quantity is not interpreted as a competitive score to be minimised globally across voices. Instead, it functions as a local measure of viability, capturing how well a given voice maintains coherence under the shared constraints imposed by the observations.

From this perspective, voices operate as semi-independent inferential processes. Each seeks to preserve its own internal consistency, yet none is privileged a priori as the unique or correct explanation of the data.

\subsection{A polyphonic free energy functional}

To formalise coordination among voices, these local free energies are combined within a polyphonic free energy functional that incorporates explicit coupling terms \citep{boyd2011admm,hinton2002poe},
\begin{equation}
F_{\mathrm{poly}} = \sum_{k=1}^K \pi_k F_k + \sum_{i<j} \lambda_{ij} C(q_i,q_j).
\end{equation}

Here, the coefficients $\pi_k$ are non-negative integration weights satisfying $\sum_k \pi_k = 1$, while $C(q_i,q_j)$ denotes a coupling functional that penalises incompatibility or promotes soft alignment between pairs of voices. The coupling strengths $\lambda_{ij}$ control the extent to which voices influence one another, allowing coordination to be tuned continuously rather than imposed categorically.

Several features of this formulation are worth emphasising: rather than being minimised independently, the individual free energies $F_k$ contribute jointly to a global objective that balances internal coherence with cross-voice compatibility. Moreover, coordination is enforced at the level of relations between voices, rather than through direct agreement at the level of latent variables. As a result, voices are able to remain distinct while still participating in a coherent collective configuration \citep{boyd2011admm,genest1986combining}.

The specific form of the coupling functional $C(q_i,q_j)$ is necessarily model-dependent. In practice, it may be defined in terms of divergences between predicted observations, overlaps between policy distributions, or measures of state compatibility. What matters for the present argument is not the particular choice, but the general principle that coordination is achieved through soft relational constraints rather than hard selection.

\subsection{Non-dominating integration and bounded influence}

For polyphonic inference to remain genuinely pluralistic, integration among voices must be non-dominating. This requirement can be expressed formally by constraining the dynamics of the integration weights $\pi_k$ and the coupling terms $\lambda_{ij}$, such that no voice is permitted to collapse to zero influence or to assume complete control.

One practical mechanism for achieving this involves bounded or leaky update rules for $\pi_k$, in which evidence accumulates gradually but is subject to lower and upper bounds \citep{rose1998deterministic,kirkpatrick1983optimization}. Under such dynamics, voices that temporarily perform poorly are not eliminated outright, preserving diversity and allowing alternative explanations to regain influence as conditions change or new data become available.

This behaviour distinguishes polyphonic inference from ensemble averaging and Bayesian model selection, where weights typically converge toward dominance as evidence accumulates \citep{hoeting1999bma,blei2017variational}. In contrast, polyphonic inference treats collapse not as a desirable outcome, but as a failure mode to be avoided.

\subsection{Relation to optimisation and stability}

Although the polyphonic framework introduces a global functional, it does not simply reintroduce classical convergence under a different name. The objective $F_{\mathrm{poly}}$ does not define a unique optimum toward which the system must converge. Instead, it specifies a constraint surface within which coordinated configurations of voices can persist.

Local optimisation may still occur within individual voices, yet global behaviour is governed by the maintenance of a viable polyphonic configuration rather than by convergence to a single solution. Viewed from a dynamical systems perspective, this replaces fixed-point convergence with a form of structured metastability, in which voices may transiently align, diverge, or reweight their influence without collapsing into a single attractor \citep{tognoli2014metastable,kelso1995dynamic,rabinovich2008neuro}. Such dynamics naturally accommodate non-stationarity, ambiguity, and open-ended learning.

Concrete instantiations of these ideas are provided in the following section. The examples that follow are intended not as performance benchmarks, but as proofs of principle demonstrating that non-dominating, pluralistic inference can be implemented in simple computational systems.

\section{Proof-of-Principle Examples}

The purpose of this section is to demonstrate that the principles of polyphonic intelligence can be instantiated in simple computational systems in a way that preserves plurality, maintains coordination without collapse, and produces behaviour that is qualitatively distinct from convergent alternatives. Rather than evaluating performance in terms of speed, accuracy, or benchmark optimisation, the focus throughout is on structural properties: whether multiple voices remain viable over time, whether coordination emerges through relational constraints rather than centralised arbitration, and whether the resulting dynamics reflect the non-dominating integration formalised in Section~4.

\subsection{Polyphonic variational inference: a toy example}

A natural point of departure is a minimal inference problem in which the posterior distribution over latent variables is explicitly multimodal, as is common in inverse modelling, perceptual inference, and system identification, yet is typically handled in practice by selecting a single dominant explanation or by averaging across alternatives in a way that suppresses their distinct structure \citep{tarantola2005inverse,kaipio2005statistical}.

In the polyphonic variational setting, multiple variational approximations are maintained in parallel, each corresponding to a distinct voice that performs local free-energy minimisation to preserve its own internal coherence, while soft coupling terms act on predicted observations to shape the compatibility relations among voices without forcing agreement in latent space. Under this formulation, no mechanism is introduced that explicitly privileges a single explanatory trajectory, and no voice is required to relinquish influence simply because another achieves a marginally lower free-energy value at a given iteration \citep{blei2017variational,wainwright2008graphical,turner2011two}.

When applied to a bimodal likelihood defined over a two-dimensional latent space, this configuration gives rise to a pattern of behaviour in which distinct explanatory trajectories persist across optimisation, with subsets of voices gravitating toward different basins of attraction while remaining weakly coordinated through shared constraints on their predicted outcomes. Coordination therefore emerges as a relational property of the ensemble rather than as the result of convergence toward a unique solution, allowing plurality to be preserved even as the system maintains global coherence.

\begin{figure}[t]
\centering
\includegraphics[width=\linewidth]{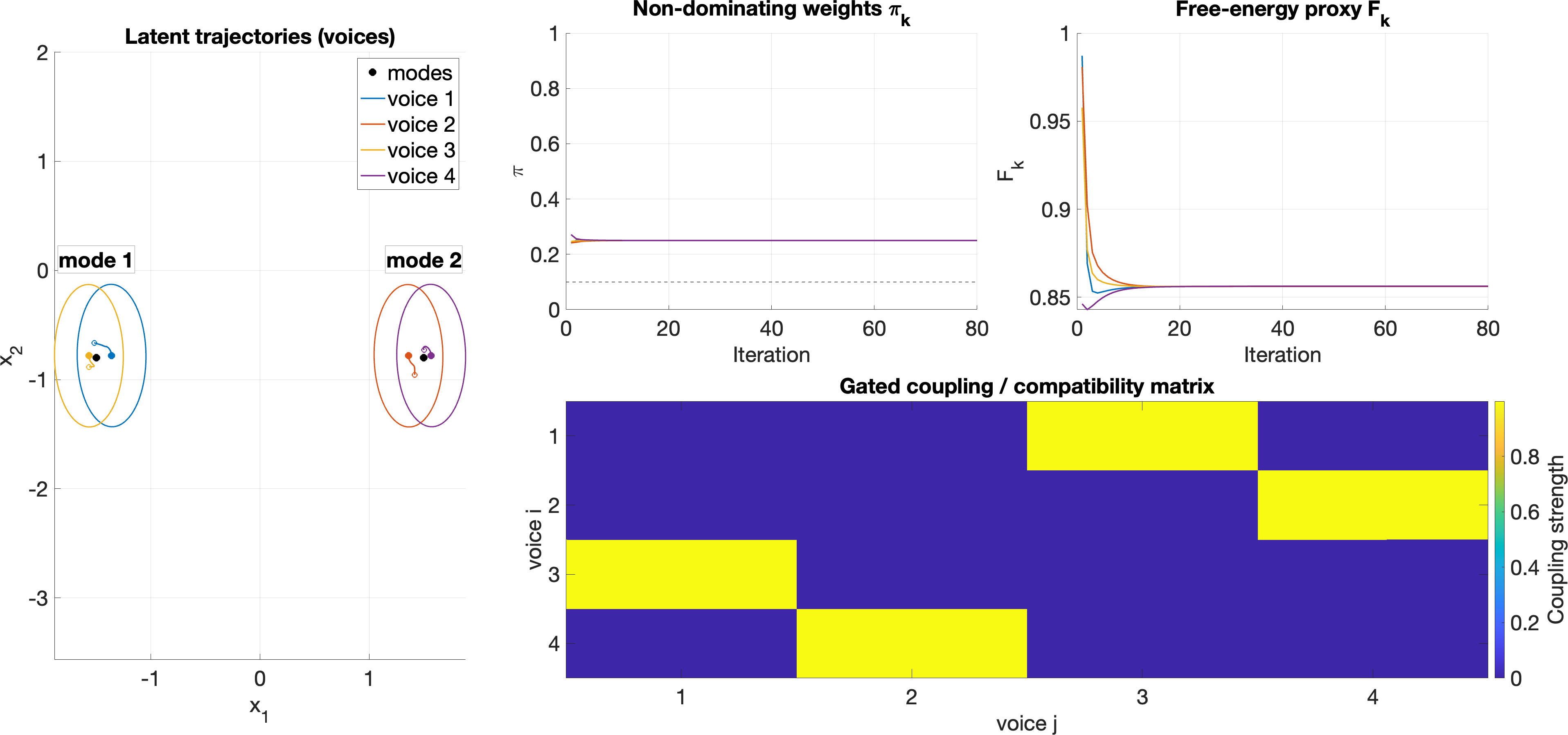}
\caption{Polyphonic variational inference in a multimodal latent space. 
Trajectories of four voices evolving under a polyphonic free-energy functional in a two-dimensional latent space with a bimodal likelihood. Final posterior covariances are shown as ellipses. Two distinct posterior modes remain occupied, with strong within-mode compatibility and weak cross-mode coupling. The evolution of non-dominating integration weights $\pi_k$ demonstrates bounded influence, preventing winner-takes-all collapse, while voice-level free-energy proxies stabilise without eliminating subdominant explanatory trajectories. The final compatibility matrix illustrates how coordination is structured by relational alignment among voices rather than by global dominance.}
\label{fig:polyphonic_vl}
\end{figure}

Viewed through this lens, Figure~\ref{fig:polyphonic_vl} illustrates how plurality is sustained rather than eliminated, how coordination is mediated primarily at the level of predicted observations rather than latent variables, and how the bounded evolution of integration weights prevents the ensemble from collapsing onto a single explanatory narrative even in the presence of small differences in local free energy. The resulting configuration is therefore better characterised as a viable plural system than as an approximation converging toward a unique posterior summary.

\subsection{Polyphonic action selection in an active inference agent}

The same organisational principles extend naturally from perceptual inference to decision-making and control. To make this concrete, we consider a minimal one-dimensional ``pong-like'' environment in which an agent controls a paddle in order to track the vertical position of a moving ball, while maintaining multiple internal policies that differ in their planning horizons, effort sensitivities, and assumptions about the ball’s future dynamics.

Rather than selecting a single policy through the global minimisation of expected free energy \citep{friston2015active,friston2017active,parr2019generalised}, action selection proceeds by integrating the action proposals of multiple voices using bounded, non-dominating weights, such that partially incompatible behavioural tendencies remain active and continue to shape the agent’s behaviour. In this setting, local optimisation occurs within each voice, but global behaviour is determined by the evolving pattern of compatibility and reweighting across the ensemble rather than by discrete policy switching or centralised arbitration \citep{daCosta2020active,parr2022active}.

When the environment undergoes an unannounced regime change, implemented here as a reversal in the ball’s velocity, the voices respond heterogeneously according to their respective generative assumptions, leading to a gradual reconfiguration of their relative influence rather than an abrupt collapse onto a newly dominant strategy. The integrated policy therefore remains coherent and adaptive, not because uncertainty has been eliminated, but because multiple inferential trajectories remain available and are re-coordinated under the new constraint structure.

\begin{figure}[t]
\centering
\includegraphics[width=\linewidth]{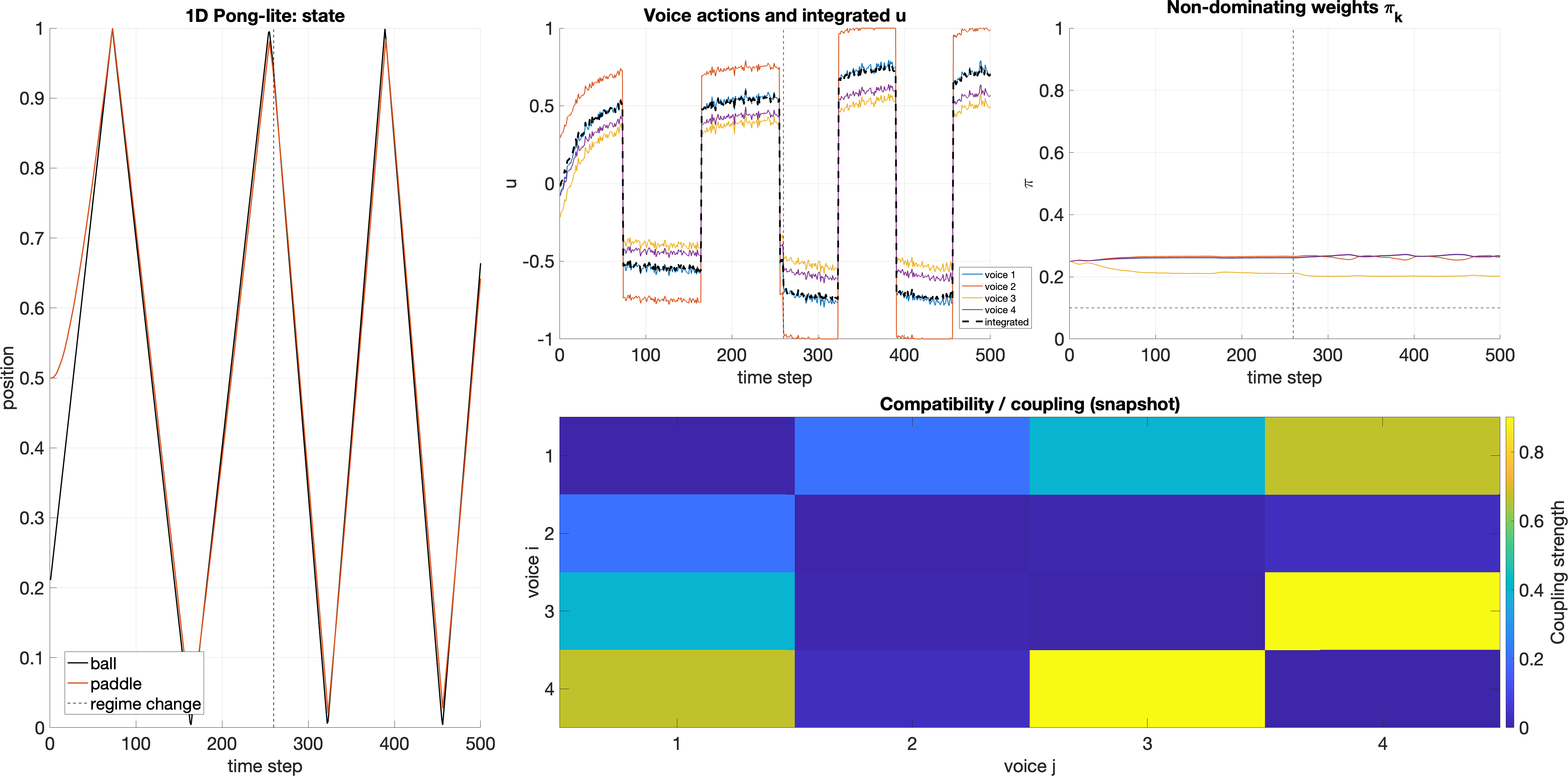}
\caption{Polyphonic action selection in a 1D Pong-like environment. 
Ball and paddle trajectories under a mid-episode velocity reversal (vertical dashed line), alongside voice-specific action proposals and the integrated polyphonic action. Non-dominating integration weights $\pi_k$ evolve smoothly in response to the regime change while remaining bounded away from collapse. The final compatibility matrix, computed from action similarity, reveals structured coordination among subsets of voices rather than global alignment or policy dominance.}
\label{fig:polyphonic_pong}
\end{figure}

As illustrated in Figure~\ref{fig:polyphonic_pong}, the resulting behaviour is most naturally understood in terms of viability rather than optimality, insofar as the system maintains coherent control by reorganising the influence of plural action tendencies rather than by suppressing disagreement or enforcing convergence on a single internal policy.

\subsection{Conceptual mapping to biological and cognitive systems}

Beyond these computational instantiations, the polyphonic framework provides a natural lens through which to interpret biological and cognitive phenomena that are characterised by persistent plurality, such as competing perceptual interpretations, overlapping neural representations, and the simultaneous availability of multiple action affordances \citep{blake2002visual,leopold1999multistable}.

From this perspective, phenomena including perceptual bistability, mixed-selectivity at the level of neural populations \citep{rigotti2013mixed,fusi2016mixed}, and heterogeneity in psychiatric symptom profiles  can be understood not as failures of convergence \citep{insel2010rdoc,fried2015depression}, but as expressions of viable configurations in which multiple inferential trajectories remain active under shared physiological, sensory, and behavioural constraints. Rather than treating such plurality as noise or inefficiency, the polyphonic view frames it as a structural feature of systems that prioritise adaptability and responsiveness in uncertain and non-stationary environments.

These correspondences are not intended as mechanistic explanations, but as conceptual alignments that situate polyphonic intelligence within a broader landscape of empirical observations, reinforcing its role as a general organisational principle rather than a task-specific modelling strategy.

For transparency and reproducibility, full MATLAB implementations of both proof-of-principle examples, including the polyphonic variational inference toy model and the polyphonic active inference Pong agent, are openly available at \url{https://github.com/alexandershaw4/polyphonic_intelligence}. The repository is intended not as a polished software package, but as a minimal, readable reference implementation that exposes the core design choices underlying voices, coupling, and non-dominating integration, and that can be readily adapted for further methodological or theoretical exploration.

\section{Implications}

Seen in light of the preceding sections, polyphonic intelligence carries implications that cut across neuroscience, psychiatry, and artificial intelligence, primarily by shifting how plurality, coordination, and adaptation are interpreted. Treating multiplicity as a stable and productive mode of organisation rather than as a problem to be resolved encourages a reframing of intelligence that emphasises structure and interaction over resolution and collapse. The emphasis in what follows is therefore interpretive rather than predictive, with the aim of clarifying how a polyphonic perspective reorients existing observations rather than proposing immediate empirical tests.

\subsection{Implications for neuroscience}

Across contemporary neuroscience, increasing attention has been paid to the distributed, redundant, and overlapping nature of neural representations, with neural populations routinely encoding multiple features simultaneously, exhibiting mixed selectivity, and participating in transient coalitions rather than occupying fixed functional roles. When viewed through models that assume rapid convergence toward a single internal representation, such properties can appear inefficient or noisy, yet they are among the most robust and reproducible features of neural organisation across species and scales \citep{pouget2000information,decharms2000neural,rigotti2013mixed,fusi2016mixed}.

Interpreted polyphonically, these observations take on a different character. Multiple neural populations, circuits, or dynamical modes can be understood as voices that maintain partially independent inferential trajectories while remaining coordinated through shared sensory, metabolic, and behavioural constraints, such that apparent redundancy reflects flexibility rather than waste. From this perspective, neural variability is not something to be averaged away, but a signature of a system organised around viability and adaptability rather than strict optimisation \citep{tognoli2014metastable,deco2012ongoing}.

This view also dovetails naturally with dynamical systems accounts that emphasise metastability, criticality, and transient synchronisation, in which neural activity does not converge to fixed-point attractors but instead occupies structured regions of state space that support multiple coexisting interpretations or action tendencies. Polyphonic intelligence provides a computational lens through which such dynamics can be understood as inferential processes unfolding under constraint \citep{tognoli2014metastable,beggs2003neuronal,deco2012ongoing}, linking population-level coordination to the sustained coexistence of plural internal states.

\subsection{Implications for psychiatry and mental health}

Within psychiatry, heterogeneity is the norm rather than the exception, with individuals sharing similar symptom profiles often differing substantially in underlying neurobiological mechanisms, and single individuals exhibiting multiple, sometimes competing, cognitive and affective tendencies across time and context. Convergent explanatory models, which seek to identify a single underlying dysfunction or optimal explanatory state, frequently struggle to accommodate this diversity, leading either to oversimplification or to increasingly fine-grained stratification imposed after the fact \citep{insel2010rdoc,fried2015depression,huys2016computational}.

A polyphonic perspective offers an alternative way of organising these observations. Rather than treating psychiatric symptoms as failures of inference per se, they can be understood as reflecting alterations in how voices are coordinated, weighted, or coupled, including the persistence of incompatible inferential trajectories, weakened constraint structures, or excessive dominance of particular voices. Such configurations may remain locally viable, particularly in the short term or within specific environments, while becoming maladaptive under other conditions \citep{borsboom2017network,huys2016computational}.

Reframing psychopathology in this way shifts emphasis away from the search for a single correct model of disorder and toward characterising patterns of coordination, rigidity, and imbalance across voices. It also opens space for intervention strategies that aim to reshape coupling and constraint structures rather than suppressing particular representations or behaviours outright. While these implications remain speculative, they illustrate how a polyphonic framework naturally accommodates heterogeneity, context dependence, and dynamical change without requiring premature resolution \citep{insel2010rdoc,borsboom2017network}.

\subsection{Implications for artificial intelligence and adaptive systems}

In artificial intelligence, ongoing efforts to build increasingly general and robust systems have exposed the limitations of strictly convergent optimisation, with models trained to optimise a single objective often exhibiting brittleness, poor generalisation, and sensitivity to distributional shift. These concerns have motivated interest in ensembles, modular architectures, and hybrid symbolic–subsymbolic systems \citep{quinonero2009dataset,goodfellow2015explaining,ovadia2019can}, although many such approaches continue to rely on eventual dominance or arbitration to resolve internal disagreement \citep{dietterich2000ensemble,jacobs1991adaptive,shazeer2017outrageously}.

Polyphonic intelligence suggests a different architectural stance, one in which multiple inferential processes remain partially autonomous while coordinating through soft constraints rather than collapsing into a single internal narrative. By preserving alternative hypotheses and action tendencies, such systems may retain the capacity to adapt when conditions change, drawing on plural internal structure rather than relying solely on stochastic exploration.

This framing also intersects with growing interest in interpretability and alignment, since maintaining multiple voices with bounded influence allows internal disagreement to remain explicit rather than hidden within aggregated representations. Instead of forcing premature resolution, uncertainty and plurality remain visible and structured, supporting more flexible and transparent forms of decision-making \citep{doshi2017towards,lipton2018mythos,amodei2016concrete}.

\subsection{Toward pluralistic computational architectures}

Taken more broadly, the polyphonic perspective points toward a class of computational architectures in which coherence arises through coordination rather than command. Rather than relying on centralised controllers or monolithic objectives, such systems consist of interacting components whose relations are shaped by shared constraints \citep{mitchell2009complex,newman2010networks} and mutual compatibility.

Within this view, optimisation is not abandoned, but repositioned as a local and contextual process operating within individual voices, while global behaviour reflects the maintenance of a viable plural configuration rather than convergence to a single solution. This shift parallels developments in complex systems science, where global order is understood to emerge from local interactions rather than top-down control \citep{mitchell2009complex,newman2010networks}.

Although substantial open questions remain concerning scalability, learning dynamics, and formal guarantees, polyphonic intelligence provides a conceptual foundation for exploring pluralistic architectures in a systematic way. In doing so, it invites a re-examination of intelligence not as the elimination of ambiguity, but as the capacity to live with it productively.

\section{Limitations and Open Questions}

In moving beyond convergent models of inference and control, polyphonic intelligence necessarily brings with it a set of open questions and limitations. These are not best understood as shortcomings of the framework, but as challenges that arise whenever pluralistic architectures are taken seriously rather than treated as transitional approximations. Similar tensions have long been recognised in distributed optimisation, complex systems, and cybernetics, where coordination among multiple interacting components introduces both new forms of robustness and new sources of instability \citep{ashby1956design,mitchell2009complex}. Clarifying these issues is essential both for theoretical development and for assessing the scope within which polyphonic organisation is likely to be most appropriate.

\subsection{Computational cost and scalability}

Maintaining multiple voices in parallel entails an increase in computational cost relative to single-model inference, since several inferential processes must be updated, coupled, and integrated simultaneously. This tension mirrors classic results in distributed optimisation and consensus algorithms, where communication and coordination overhead grow with the number of agents or subsystems involved \citep{boyd2011admm,olfati2007consensus}. While such costs may be manageable in low-dimensional or modular settings, questions remain about how polyphonic architectures scale to large models and high-dimensional data, particularly in regimes that demand real-time inference or control.

At the same time, the prevalence of massive parallelism in biological systems suggests that computational cost alone need not represent a fundamental barrier. Hierarchical organisation, sparse coupling, and selective activation are widely observed strategies for managing complexity in both neural and engineered systems \citep{newman2010networks,mitchell2009complex}. Polyphonic architectures may admit analogous approximations, including dynamically recruiting subsets of voices or modulating coupling strength in response to task demands, yet the conditions under which such strategies preserve plurality without eroding coherence remain largely unexplored.

\subsection{Stability, learning dynamics, and failure modes}

Although non-dominating integration is central to the polyphonic framework, it also raises questions about stability and long-term learning dynamics. In coupled dynamical systems, weak interactions can lead to fragmentation and loss of global coordination, while overly strong coupling can induce synchronisation or collapse into low-dimensional attractors \citep{strogatz2001exploring,pikovsky2003synchronization}. Polyphonic systems sit deliberately between these regimes, yet the boundaries of this intermediate zone are not well characterised.

Understanding how voices should be introduced, modified, or retired over time, and how coupling strengths should evolve as evidence accumulates, connects naturally to broader work on metastability and bifurcation structure in neural and cognitive systems \citep{tognoli2014metastable,deco2012ongoing}. Formal analyses of these dynamics, including the identification of failure modes such as runaway dominance, oscillatory instability, or incoherent pluralism, remain an open and important direction for future research.

\subsection{Relation to existing pluralistic and ensemble methods}

Although polyphonic intelligence is conceptually distinct from ensemble learning, mixture models, and Bayesian model averaging, the boundaries between these approaches are not always sharply defined in practice. Under certain choices of coupling functions or update rules, polyphonic implementations may approximate weighted ensembles or structured mixtures, particularly when integration weights evolve slowly or compatibility constraints are weak \citep{dietterich2000ensemble,hoeting1999bma,mclachlan2000finite}.

Disentangling genuine non-dominating integration from softened forms of convergence therefore presents both a theoretical and an empirical challenge. One possible route is to characterise polyphonic systems in terms of dynamical signatures, such as the persistence of multiple metastable configurations or bounded weight trajectories, rather than solely in terms of architectural form \citep{tognoli2014metastable}. Developing such criteria would help distinguish coordination through sustained plurality from systems that merely delay or smooth the path to dominance.

\subsection{Interpretability and evaluation}

Assessing the performance of polyphonic systems raises challenges that differ from those encountered in convergent models. Standard metrics such as predictive accuracy, loss minimisation, or regret are well suited to single-objective optimisation, but they may fail to capture properties such as diversity, resilience to distributional shift, or the capacity to reorganise under uncertainty \citep{ovadia2019can,quinonero2009dataset}. As a result, alternative criteria may be required that explicitly assess coordination, heterogeneity, and responsiveness rather than final performance alone.

At the same time, while maintaining multiple voices has the potential to enhance interpretability by making internal disagreement explicit, it also complicates analysis by introducing interacting explanatory threads. This tension echoes broader debates in interpretable machine learning, where transparency and complexity often stand in opposition \citep{doshi2017towards,lipton2018mythos}. Developing tools that allow polyphonic internal states to be visualised, summarised, and interrogated without collapsing them into a single narrative remains an important methodological challenge.

\subsection{Scope and normative assumptions}

Finally, it is important to emphasise that polyphonic intelligence is not intended as a universal replacement for convergent models. In well-defined, stationary, and low-dimensional settings, classical optimisation and single-model inference often remain both efficient and appropriate, as evidenced by their continued success across signal processing and control applications \citep{boyd2004convex,kalman1960new}. The framework instead challenges the assumption that convergence should be treated as the default or defining property of intelligence across all domains.

Determining when polyphonic organisation is beneficial, and when it is unnecessary or even counterproductive, will require careful empirical and theoretical investigation. This includes identifying task characteristics, environmental conditions, and system constraints that favour pluralistic coordination over singular optimisation, as well as clarifying the normative commitments that underlie different modelling choices \citep{wolpert1997no,ashby1956design}.

\section{Conclusion}

Throughout this paper, we have developed \emph{polyphonic intelligence} as an alternative framing for understanding intelligent behaviour in both biological and artificial systems. Rather than treating convergence toward a single objective, representation, or policy as a defining feature of intelligence, the polyphonic perspective emphasises coordination among multiple semi-independent processes operating under shared constraints. Within this view, coherence arises not through dominance or centralised control, but through soft alignment, compatibility, and the sustained coexistence of plural inferential trajectories, echoing long-standing themes in coordination dynamics and complex systems theory \citep{kelso1995dynamic,mitchell2009complex}.

By tracing how convergence has become embedded in dominant models of intelligence for reasons of tractability and engineering convenience, we argued that many existing attempts to accommodate multiplicity ultimately reintroduce dominance at the level of integration. Polyphonic intelligence addresses this gap by treating plurality itself as a stable and productive mode of organisation, formalised through non-dominating integration within a variational framework. In doing so, it extends the logic of variational free energy and distributed optimisation toward a regime in which coordination is relational rather than eliminative \citep{jordan1999variational,boyd2011admm}.

Grounding the framework in established principles from variational inference, dynamical systems theory, and control ensures compatibility with existing formalisms while simultaneously challenging their default assumptions. The proof-of-principle examples demonstrate that pluralistic, non-convergent inference is not only conceptually coherent but computationally well defined, and that its behaviour differs qualitatively from ensemble and mixture-based approaches even in simple settings, particularly in its ability to preserve metastable configurations rather than collapse toward a single explanatory state \citep{tognoli2014metastable,dietterich2000ensemble}.

Viewed more broadly, polyphonic intelligence invites a shift in how intelligence is evaluated and designed. Rather than prioritising optimality and certainty as primary goals, it foregrounds viability, adaptability, and coordination under constraint, aligning naturally with perspectives from computational neuroscience and psychiatry that emphasise heterogeneity, context dependence, and dynamical organisation over fixed explanatory categories \citep{huys2016computational,insel2010rdoc}.

In reframing intelligence as coordination without command, the polyphonic perspective does not reject optimisation outright, but situates it within a richer organisational landscape in which local optimisation serves plural global configurations rather than dictating them. How far this framing can be pushed remains an open empirical and theoretical question. Nevertheless, by making plurality explicit rather than suppressing it, polyphonic intelligence offers a conceptual lens through which complex adaptive behaviour can be understood not as a failure to converge, but as an achievement in its own right, consistent with broader principles of viability and self-organisation in adaptive systems \citep{aubin1991viability,ashby1956design}.

\bibliographystyle{plainnat}
\bibliography{refs}

\begin{thebibliography}{84}
\providecommand{\natexlab}[1]{#1}
\providecommand{\url}[1]{\texttt{#1}}
\expandafter\ifx\csname urlstyle\endcsname\relax
  \providecommand{\doi}[1]{doi: #1}\else
  \providecommand{\doi}{doi: \begingroup \urlstyle{rm}\Url}\fi

\bibitem[Amodei et~al.(2016)Amodei, Olah, Steinhardt, Christiano, Schulman, and Man{\'e}]{amodei2016concrete}
Dario Amodei, Chris Olah, Jacob Steinhardt, Paul Christiano, John Schulman, and Dan Man{\'e}.
\newblock Concrete problems in {AI} safety.
\newblock \emph{arXiv preprint arXiv:1606.06565}, 2016.

\bibitem[Ashby(1956)]{ashby1956design}
W.~Ross Ashby.
\newblock \emph{An Introduction to Cybernetics}.
\newblock Chapman \& Hall, London, 1956.

\bibitem[Aubin(1991)]{aubin1991viability}
Jean-Pierre Aubin.
\newblock \emph{Viability Theory}.
\newblock Birkh{\"a}user, 1991.

\bibitem[Barandiaran et~al.(2009)Barandiaran, Di~Paolo, and Rohde]{barandiaran2009defining}
Xabier~E. Barandiaran, Ezequiel Di~Paolo, and Marieke Rohde.
\newblock Defining agency: Individuality, normativity, asymmetry, and spatiotemporality in action.
\newblock \emph{Adaptive Behavior}, 17\penalty0 (5):\penalty0 367--386, 2009.

\bibitem[Beer(2000)]{beer2000dynamical}
Randall~D. Beer.
\newblock Dynamical approaches to cognitive science.
\newblock \emph{Trends in Cognitive Sciences}, 4\penalty0 (3):\penalty0 91--99, 2000.

\bibitem[Beggs and Plenz(2003)]{beggs2003neuronal}
John~M. Beggs and Dietmar Plenz.
\newblock Neuronal avalanches in neocortical circuits.
\newblock \emph{Journal of Neuroscience}, 23\penalty0 (35):\penalty0 11167--11177, 2003.

\bibitem[Bertsekas(2017)]{bertsekas2017dp}
Dimitri~P. Bertsekas.
\newblock \emph{Dynamic Programming and Optimal Control}.
\newblock Athena Scientific, Belmont, MA, 4 edition, 2017.

\bibitem[Bishop(2006)]{bishop2006pattern}
Christopher~M. Bishop.
\newblock \emph{Pattern Recognition and Machine Learning}.
\newblock Springer, 2006.

\bibitem[Blake and Logothetis(2002)]{blake2002visual}
Randolph Blake and Nikos~K. Logothetis.
\newblock Visual competition.
\newblock \emph{Nature Reviews Neuroscience}, 3\penalty0 (1):\penalty0 13--21, 2002.

\bibitem[Blei et~al.(2017)Blei, Kucukelbir, and McAuliffe]{blei2017variational}
David~M. Blei, Alp Kucukelbir, and Jon~D. McAuliffe.
\newblock Variational inference: A review for statisticians.
\newblock \emph{Journal of the American Statistical Association}, 112\penalty0 (518):\penalty0 859--877, 2017.

\bibitem[Bogacz et~al.(2006)Bogacz, Brown, Moehlis, Holmes, and Cohen]{bogacz2006physics}
Rafal Bogacz, Eric Brown, Jeff Moehlis, Philip Holmes, and Jonathan~D. Cohen.
\newblock The physics of optimal decision making: A formal analysis of models of performance in two-alternative forced-choice tasks.
\newblock \emph{Psychological Review}, 113\penalty0 (4):\penalty0 700--765, 2006.

\bibitem[Borsboom(2017)]{borsboom2017network}
Denny Borsboom.
\newblock A network theory of mental disorders.
\newblock \emph{World Psychiatry}, 16\penalty0 (1):\penalty0 5--13, 2017.

\bibitem[Boyd and Vandenberghe(2004)]{boyd2004convex}
Stephen Boyd and Lieven Vandenberghe.
\newblock \emph{Convex Optimization}.
\newblock Cambridge University Press, Cambridge, UK, 2004.

\bibitem[Boyd et~al.(2011)Boyd, Parikh, Chu, Peleato, and Eckstein]{boyd2011admm}
Stephen Boyd, Neal Parikh, Eric Chu, Borja Peleato, and Jonathan Eckstein.
\newblock Distributed optimization and statistical learning via the alternating direction method of multipliers.
\newblock \emph{Foundations and Trends{\textregistered} in Machine Learning}, 3\penalty0 (1):\penalty0 1--122, 2011.

\bibitem[Breakspear(2017)]{breakspear2017dynamic}
Michael Breakspear.
\newblock Dynamic models of large-scale brain activity.
\newblock \emph{Nature Neuroscience}, 20:\penalty0 340--352, 2017.

\bibitem[Cisek(2007)]{cisek2007cortical}
Paul Cisek.
\newblock Cortical mechanisms of action selection: The affordance competition hypothesis.
\newblock \emph{Philosophical Transactions of the Royal Society B}, 362\penalty0 (1485):\penalty0 1585--1599, 2007.

\bibitem[Clark(2013)]{clark2013whatever}
Andy Clark.
\newblock Whatever next? predictive brains, situated agents, and the future of cognitive science.
\newblock \emph{Behavioral and Brain Sciences}, 36\penalty0 (3):\penalty0 181--204, 2013.

\bibitem[Couzin et~al.(2005)Couzin, Krause, Franks, and Levin]{couzin2005effective}
Iain~D. Couzin, Jens Krause, Nigel~R. Franks, and Simon~A. Levin.
\newblock Effective leadership and decision-making in animal groups on the move.
\newblock \emph{Nature}, 433:\penalty0 513--516, 2005.

\bibitem[da~Costa et~al.(2020)da~Costa, Parr, Sajid, Veselic, Neacsu, and Friston]{daCosta2020active}
Lancelot da~Costa, Thomas Parr, Noor Sajid, Senka Veselic, Valentin Neacsu, and Karl Friston.
\newblock Active inference on discrete state-spaces: A synthesis.
\newblock \emph{Journal of Mathematical Psychology}, 99:\penalty0 102447, 2020.

\bibitem[Dayan and Abbott(2001)]{dayan2001theoretical}
Peter Dayan and L.F. Abbott.
\newblock \emph{Theoretical Neuroscience: Computational and Mathematical Modeling of Neural Systems}.
\newblock MIT Press, Cambridge, MA, 2001.

\bibitem[deCharms and Zador(2000)]{decharms2000neural}
R.~Christopher deCharms and Anthony Zador.
\newblock Neural representation and the cortical code.
\newblock \emph{Annual Review of Neuroscience}, 23:\penalty0 613--647, 2000.

\bibitem[Deco and Jirsa(2012)]{deco2012ongoing}
Gustavo Deco and Viktor~K. Jirsa.
\newblock Ongoing cortical activity at rest: Criticality, multistability, and ghost attractors.
\newblock \emph{The Neuroscientist}, 18\penalty0 (5):\penalty0 523--535, 2012.

\bibitem[Di~Paolo et~al.(2010)Di~Paolo, Rohde, and De~Jaegher]{dipaolo2010enactive}
Ezequiel~A. Di~Paolo, Marieke Rohde, and Hanne De~Jaegher.
\newblock Horizons for the enactive mind: Values, social interaction, and play.
\newblock \emph{Enactive Mind}, 1:\penalty0 33--87, 2010.

\bibitem[Dietterich(2000)]{dietterich2000ensemble}
Thomas~G. Dietterich.
\newblock Ensemble methods in machine learning.
\newblock In \emph{Multiple Classifier Systems}, pages 1--15. Springer, 2000.

\bibitem[Doshi-Velez and Kim(2017)]{doshi2017towards}
Finale Doshi-Velez and Been Kim.
\newblock Towards a rigorous science of interpretable machine learning.
\newblock \emph{arXiv preprint arXiv:1702.08608}, 2017.

\bibitem[Fried and Nesse(2015)]{fried2015depression}
Eiko~I. Fried and Randolph~M. Nesse.
\newblock Depression is not a consistent syndrome: An investigation of unique symptom patterns in the {STAR*D} study.
\newblock \emph{Journal of Affective Disorders}, 172:\penalty0 96--102, 2015.

\bibitem[Friston(2010)]{friston2010free}
Karl Friston.
\newblock The free-energy principle: a unified brain theory?
\newblock \emph{Nature Reviews Neuroscience}, 11\penalty0 (2):\penalty0 127--138, 2010.

\bibitem[Friston et~al.(2015)Friston, Rigoli, Ognibene, Mathys, Fitzgerald, and Pezzulo]{friston2015active}
Karl Friston, Francesco Rigoli, Dimitri Ognibene, Christoph Mathys, Thomas Fitzgerald, and Giovanni Pezzulo.
\newblock Active inference and epistemic value.
\newblock \emph{Cognitive Neuroscience}, 6\penalty0 (4):\penalty0 187--214, 2015.

\bibitem[Friston et~al.(2017)Friston, FitzGerald, Rigoli, Schwartenbeck, and Pezzulo]{friston2017active}
Karl Friston, Thomas FitzGerald, Francesco Rigoli, Philipp Schwartenbeck, and Giovanni Pezzulo.
\newblock Active inference: A process theory.
\newblock \emph{Neural Computation}, 29\penalty0 (1):\penalty0 1--49, 2017.

\bibitem[Fusi et~al.(2016)Fusi, Miller, and Rigotti]{fusi2016mixed}
Stefano Fusi, Earl~K. Miller, and Mattia Rigotti.
\newblock Why neurons mix: high dimensionality for higher cognition.
\newblock \emph{Current Opinion in Neurobiology}, 37:\penalty0 66--74, 2016.

\bibitem[Gelman et~al.(2013)Gelman, Carlin, Stern, Dunson, Vehtari, and Rubin]{gelman2013bda}
Andrew Gelman, John~B. Carlin, Hal~S. Stern, David~B. Dunson, Aki Vehtari, and Donald~B. Rubin.
\newblock \emph{Bayesian Data Analysis}.
\newblock CRC Press, Boca Raton, FL, 3 edition, 2013.

\bibitem[Genest and Zidek(1986)]{genest1986combining}
Christian Genest and James~V. Zidek.
\newblock Combining probability distributions: A critique and an annotated bibliography.
\newblock \emph{Statistical Science}, 1\penalty0 (1):\penalty0 114--148, 1986.

\bibitem[Gershman et~al.(2014)Gershman, Hoffman, and Blei]{gershman2012nonparametric}
Samuel~J. Gershman, Matthew~D. Hoffman, and David~M. Blei.
\newblock Nonparametric variational inference.
\newblock \emph{Journal of Machine Learning Research}, 15:\penalty0 1--29, 2014.

\bibitem[Goodfellow et~al.(2015)Goodfellow, Shlens, and Szegedy]{goodfellow2015explaining}
Ian~J. Goodfellow, Jonathon Shlens, and Christian Szegedy.
\newblock Explaining and harnessing adversarial examples.
\newblock In \emph{International Conference on Learning Representations (ICLR)}, 2015.

\bibitem[Gutenkunst et~al.(2007)Gutenkunst, Waterfall, Casey, Brown, Myers, and Sethna]{gutenkunst2007universally}
Ryan~N. Gutenkunst, Joshua~J. Waterfall, F.~Casey, Kevin~S. Brown, Charles~R. Myers, and James~P. Sethna.
\newblock Universally sloppy parameter sensitivities in systems biology models.
\newblock \emph{PLoS Computational Biology}, 3\penalty0 (10):\penalty0 e189, 2007.

\bibitem[Haken(1983)]{haken1983synergetics}
Hermann Haken.
\newblock \emph{Synergetics: An Introduction}.
\newblock Springer, Berlin, 1983.

\bibitem[Hastie et~al.(2009)Hastie, Tibshirani, and Friedman]{hastie2009elements}
Trevor Hastie, Robert Tibshirani, and Jerome Friedman.
\newblock \emph{The Elements of Statistical Learning: Data Mining, Inference, and Prediction}.
\newblock Springer, New York, NY, 2 edition, 2009.

\bibitem[Hinton et~al.(2015)Hinton, Vinyals, and Dean]{hinton2015distilling}
Geoffrey Hinton, Oriol Vinyals, and Jeff Dean.
\newblock Distilling the knowledge in a neural network.
\newblock \emph{arXiv preprint arXiv:1503.02531}, 2015.

\bibitem[Hinton(2002)]{hinton2002poe}
Geoffrey~E. Hinton.
\newblock Training products of experts by minimizing contrastive divergence.
\newblock In \emph{Neural Computation}, volume~14, pages 1771--1800, 2002.

\bibitem[Hoeting et~al.(1999)Hoeting, Madigan, Raftery, and Volinsky]{hoeting1999bma}
Jennifer~A. Hoeting, David Madigan, Adrian~E. Raftery, and Chris~T. Volinsky.
\newblock Bayesian model averaging: A tutorial.
\newblock \emph{Statistical Science}, 14\penalty0 (4):\penalty0 382--417, 1999.

\bibitem[Huys et~al.(2016)Huys, Maia, and Frank]{huys2016computational}
Quentin J.~M. Huys, Tiago~V. Maia, and Michael~J. Frank.
\newblock Computational psychiatry as a bridge from neuroscience to clinical applications.
\newblock \emph{Nature Neuroscience}, 19\penalty0 (3):\penalty0 404--413, 2016.

\bibitem[Insel et~al.(2010)Insel, Cuthbert, Garvey, Heinssen, Pine, Quinn, Sanislow, and Wang]{insel2010rdoc}
Thomas Insel, Bruce Cuthbert, Michael Garvey, Robert Heinssen, Daniel~S. Pine, Kevin Quinn, Charles Sanislow, and Philip Wang.
\newblock Research domain criteria ({RDoC}): Toward a new classification framework for research on mental disorders.
\newblock \emph{American Journal of Psychiatry}, 167\penalty0 (7):\penalty0 748--751, 2010.

\bibitem[Jacobs et~al.(1991)Jacobs, Jordan, Nowlan, and Hinton]{jacobs1991adaptive}
Robert~A. Jacobs, Michael~I. Jordan, Steven~J. Nowlan, and Geoffrey~E. Hinton.
\newblock Adaptive mixtures of local experts.
\newblock \emph{Neural Computation}, 3\penalty0 (1):\penalty0 79--87, 1991.

\bibitem[Jordan et~al.(1999)Jordan, Ghahramani, Jaakkola, and Saul]{jordan1999variational}
Michael~I. Jordan, Zoubin Ghahramani, Tommi~S. Jaakkola, and Lawrence~K. Saul.
\newblock An introduction to variational methods for graphical models.
\newblock \emph{Machine Learning}, 37\penalty0 (2):\penalty0 183--233, 1999.

\bibitem[Kaelbling et~al.(1998)Kaelbling, Littman, and Cassandra]{kaelbling1998planning}
Leslie~Pack Kaelbling, Michael~L. Littman, and Anthony~R. Cassandra.
\newblock Planning and acting in partially observable stochastic domains.
\newblock \emph{Artificial Intelligence}, 101\penalty0 (1--2):\penalty0 99--134, 1998.

\bibitem[Kaipio and Somersalo(2005)]{kaipio2005statistical}
Jari Kaipio and Erkki Somersalo.
\newblock \emph{Statistical and Computational Inverse Problems}.
\newblock Springer, New York, NY, 2005.

\bibitem[Kalman(1960)]{kalman1960new}
Rudolf~E. Kalman.
\newblock A new approach to linear filtering and prediction problems.
\newblock \emph{Journal of Basic Engineering}, 82\penalty0 (1):\penalty0 35--45, 1960.

\bibitem[Kelso(1995)]{kelso1995dynamic}
J.~A.~Scott Kelso.
\newblock \emph{Dynamic Patterns: The Self-Organization of Brain and Behavior}.
\newblock MIT Press, 1995.

\bibitem[Kirkpatrick et~al.(1983)Kirkpatrick, Gelatt, and Vecchi]{kirkpatrick1983optimization}
Scott Kirkpatrick, C.~Daniel Gelatt, and Mario~P. Vecchi.
\newblock Optimization by simulated annealing.
\newblock \emph{Science}, 220\penalty0 (4598):\penalty0 671--680, 1983.

\bibitem[Lake et~al.(2017)Lake, Ullman, Tenenbaum, and Gershman]{lake2017building}
Brenden~M. Lake, Tomer~D. Ullman, Joshua~B. Tenenbaum, and Samuel~J. Gershman.
\newblock Building machines that learn and think like people.
\newblock \emph{Behavioral and Brain Sciences}, 40, 2017.

\bibitem[Leopold and Logothetis(1999)]{leopold1999multistable}
David~A. Leopold and Nikos~K. Logothetis.
\newblock Multistable perception: Isolating sensory from decision processes.
\newblock \emph{Trends in Cognitive Sciences}, 3\penalty0 (7):\penalty0 254--264, 1999.

\bibitem[Lipton(2018)]{lipton2018mythos}
Zachary~C. Lipton.
\newblock The mythos of model interpretability.
\newblock \emph{Communications of the ACM}, 61\penalty0 (10):\penalty0 36--43, 2018.

\bibitem[MacKay(2003)]{mackay2003information}
David J.~C. MacKay.
\newblock \emph{Information Theory, Inference, and Learning Algorithms}.
\newblock Cambridge University Press, 2003.

\bibitem[McLachlan and Peel(2000)]{mclachlan2000finite}
Geoffrey McLachlan and David Peel.
\newblock \emph{Finite Mixture Models}.
\newblock Wiley, New York, NY, 2000.

\bibitem[Mitchell(2009)]{mitchell2009complex}
Melanie Mitchell.
\newblock \emph{Complexity: A Guided Tour}.
\newblock Oxford University Press, Oxford, UK, 2009.

\bibitem[Neal and Hinton(1998)]{neal1998view}
Radford~M. Neal and Geoffrey~E. Hinton.
\newblock A view of the {EM} algorithm that justifies incremental, sparse, and other variants.
\newblock In Michael~I. Jordan, editor, \emph{Learning in Graphical Models}, pages 355--368. Springer, Dordrecht, 1998.

\bibitem[Newman(2010)]{newman2010networks}
Mark Newman.
\newblock \emph{Networks: An Introduction}.
\newblock Oxford University Press, Oxford, UK, 2010.

\bibitem[Olfati-Saber et~al.(2007)Olfati-Saber, Fax, and Murray]{olfati2007consensus}
Reza Olfati-Saber, J.~Alex Fax, and Richard~M. Murray.
\newblock Consensus and cooperation in networked multi-agent systems.
\newblock \emph{Proceedings of the IEEE}, 95\penalty0 (1):\penalty0 215--233, 2007.

\bibitem[Ovadia et~al.(2019)Ovadia, Fertig, Ren, Nado, Sculley, Nowozin, Dillon, Lakshminarayanan, and Snoek]{ovadia2019can}
Yaniv Ovadia, Emily Fertig, Jie Ren, Zachary Nado, D.~Sculley, Sebastian Nowozin, Joshua~V. Dillon, Balaji Lakshminarayanan, and Jasper Snoek.
\newblock Can you trust your model's uncertainty? evaluating predictive uncertainty under dataset shift.
\newblock \emph{Advances in Neural Information Processing Systems}, 32, 2019.

\bibitem[Parr and Friston(2019)]{parr2019generalised}
Thomas Parr and Karl~J. Friston.
\newblock Generalised free energy and active inference.
\newblock \emph{Biological Cybernetics}, 113\penalty0 (5--6):\penalty0 495--513, 2019.

\bibitem[Parr et~al.(2022)Parr, Pezzulo, and Friston]{parr2022active}
Thomas Parr, Giovanni Pezzulo, and Karl~J. Friston.
\newblock \emph{Active Inference: The Free Energy Principle in Mind, Brain, and Behavior}.
\newblock MIT Press, Cambridge, MA, 2022.

\bibitem[Pikovsky et~al.(2003)Pikovsky, Rosenblum, and Kurths]{pikovsky2003synchronization}
Arkady Pikovsky, Michael Rosenblum, and J{\"u}rgen Kurths.
\newblock \emph{Synchronization: A Universal Concept in Nonlinear Sciences}.
\newblock Cambridge University Press, Cambridge, UK, 2003.

\bibitem[Pouget et~al.(2000)Pouget, Dayan, and Zemel]{pouget2000information}
Alexandre Pouget, Peter Dayan, and Richard~S. Zemel.
\newblock Information processing with population codes.
\newblock \emph{Nature Reviews Neuroscience}, 1\penalty0 (2):\penalty0 125--132, 2000.

\bibitem[Qui{\~n}onero-Candela et~al.(2009)Qui{\~n}onero-Candela, Sugiyama, Schwaighofer, and Lawrence]{quinonero2009dataset}
Joaqu{\'i}n Qui{\~n}onero-Candela, Masashi Sugiyama, Anton Schwaighofer, and Neil~D. Lawrence.
\newblock Dataset shift in machine learning.
\newblock In \emph{Dataset Shift in Machine Learning}, pages 3--28. MIT Press, Cambridge, MA, 2009.

\bibitem[Rabinovich et~al.(2008)Rabinovich, Huerta, and Laurent]{rabinovich2008neuro}
Mikhail~I. Rabinovich, Ram{\'o}n Huerta, and Gilles Laurent.
\newblock Neurodynamics of attentional control: A dynamical systems perspective.
\newblock \emph{Current Opinion in Neurobiology}, 18\penalty0 (6):\penalty0 724--730, 2008.

\bibitem[Rao and Ballard(1999)]{rao1999predictive}
Rajesh P.~N. Rao and Dana~H. Ballard.
\newblock Predictive coding in the visual cortex: A functional interpretation of some extra-classical receptive-field effects.
\newblock \emph{Nature Neuroscience}, 2\penalty0 (1):\penalty0 79--87, 1999.

\bibitem[Raue et~al.(2009)Raue, Becker, Klingm{\"u}ller, and Timmer]{raue2009structural}
Andreas Raue, Philipp Becker, Ursula Klingm{\"u}ller, and Jens Timmer.
\newblock Structural and practical identifiability analysis of partially observed dynamical models by exploiting the profile likelihood.
\newblock \emph{Bioinformatics}, 25\penalty0 (15):\penalty0 1923--1929, 2009.

\bibitem[Rigotti et~al.(2013)Rigotti, Barak, Warden, Wang, Daw, Miller, and Fusi]{rigotti2013mixed}
Mattia Rigotti, Omri Barak, Melissa~R. Warden, Xiao-Jing Wang, Nathaniel~D. Daw, Earl~K. Miller, and Stefano Fusi.
\newblock The importance of mixed selectivity in complex cognitive tasks.
\newblock \emph{Nature}, 497\penalty0 (7451):\penalty0 585--590, 2013.

\bibitem[Rose(1998)]{rose1998deterministic}
Kenneth Rose.
\newblock Deterministic annealing for clustering, compression, classification, regression, and related optimization problems.
\newblock \emph{Proceedings of the IEEE}, 86\penalty0 (11):\penalty0 2210--2239, 1998.

\bibitem[Russell and Norvig(2021)]{russell2021artificial}
Stuart~J. Russell and Peter Norvig.
\newblock \emph{Artificial Intelligence: A Modern Approach}.
\newblock Pearson, Hoboken, NJ, 4 edition, 2021.

\bibitem[Shazeer et~al.(2017)Shazeer, Mirhoseini, Maziarz, Davis, Le, and Hinton]{shazeer2017outrageously}
Noam Shazeer, Azalia Mirhoseini, Krzysztof Maziarz, Andy Davis, Quoc Le, and Geoffrey Hinton.
\newblock Outrageously large neural networks: The sparsely-gated mixture-of-experts layer.
\newblock In \emph{International Conference on Learning Representations (ICLR)}, 2017.

\bibitem[Spivey(2007)]{spivey2007continuity}
Michael~J. Spivey.
\newblock \emph{The Continuity of Mind}.
\newblock Oxford University Press, Oxford, UK, 2007.

\bibitem[Sporns(2011)]{sporns2011networks}
Olaf Sporns.
\newblock \emph{Networks of the Brain}.
\newblock MIT Press, 2011.

\bibitem[Strogatz(2001)]{strogatz2001exploring}
Steven~H. Strogatz.
\newblock Exploring complex networks.
\newblock \emph{Nature}, 410\penalty0 (6825):\penalty0 268--276, 2001.
\newblock \doi{10.1038/35065725}.

\bibitem[Strogatz(2003)]{strogatz2003sync}
Steven~H. Strogatz.
\newblock \emph{Sync: The Emerging Science of Spontaneous Order}.
\newblock Hyperion, 2003.

\bibitem[Sutton and Barto(2018)]{Sutton1998}
Richard~S. Sutton and Andrew~G. Barto.
\newblock \emph{Reinforcement Learning: An Introduction}.
\newblock The MIT Press, second edition, 2018.
\newblock URL \url{http://incompleteideas.net/book/the-book-2nd.html}.

\bibitem[Tarantola(2005)]{tarantola2005inverse}
Albert Tarantola.
\newblock \emph{Inverse Problem Theory and Methods for Model Parameter Estimation}.
\newblock SIAM, Philadelphia, PA, 2005.

\bibitem[Todorov and Jordan(2002)]{todorov2002optimal}
Emanuel Todorov and Michael~I. Jordan.
\newblock Optimal feedback control as a theory of motor coordination.
\newblock \emph{Nature Neuroscience}, 5\penalty0 (11):\penalty0 1226--1235, 2002.

\bibitem[Tognoli and Kelso(2014)]{tognoli2014metastable}
Emmanuelle Tognoli and J.~A.~Scott Kelso.
\newblock The metastable brain.
\newblock \emph{Neuron}, 81\penalty0 (1):\penalty0 35--48, 2014.

\bibitem[Tononi et~al.(1994)Tononi, Sporns, and Edelman]{tononi1994measure}
Giulio Tononi, Olaf Sporns, and Gerald~M. Edelman.
\newblock A measure for brain complexity: relating functional segregation and integration in the nervous system.
\newblock \emph{Proceedings of the National Academy of Sciences}, 91\penalty0 (11):\penalty0 5033--5037, 1994.

\bibitem[Turner and Sahani(2011)]{turner2011two}
Richard~E. Turner and Maneesh Sahani.
\newblock Two problems with variational expectation maximisation for time-series models.
\newblock \emph{Bayesian Time Series Models}, pages 109--130, 2011.

\bibitem[Varela et~al.(1991)Varela, Thompson, and Rosch]{varela1991embodied}
Francisco~J. Varela, Evan Thompson, and Eleanor Rosch.
\newblock \emph{The Embodied Mind: Cognitive Science and Human Experience}.
\newblock MIT Press, Cambridge, MA, 1991.

\bibitem[Wainwright and Jordan(2008)]{wainwright2008graphical}
Martin~J. Wainwright and Michael~I. Jordan.
\newblock Graphical models, exponential families, and variational inference.
\newblock In \emph{Foundations and Trends{\textregistered} in Machine Learning}, volume~1, pages 1--305. Now Publishers, 2008.

\bibitem[Wolpert and Macready(1997)]{wolpert1997no}
David~H. Wolpert and William~G. Macready.
\newblock No free lunch theorems for optimization.
\newblock \emph{IEEE Transactions on Evolutionary Computation}, 1\penalty0 (1):\penalty0 67--82, 1997.

\end{thebibliography}

\appendix

\section{Supplementary}


\subsection*{Conventional Active Inference (single generative model)}

\paragraph{Generative model.}
Let $o_{1:T}$ denote observations, $s_{1:T}$ hidden states, and $a_{1:T}$ actions.
Under a single generative model $m$, we assume a joint density
\begin{equation}
p(o_{1:T}, s_{1:T} \mid m)
= p(s_1 \mid m)\prod_{t=1}^T p(o_t \mid s_t, m)\prod_{t=1}^{T-1} p(s_{t+1}\mid s_t, a_t, m).
\end{equation}
\noindent
This specifies (i) a prior over initial states, (ii) an observation model (likelihood),
and (iii) controlled dynamics.

\paragraph{Variational free energy (VFE).}
Active Inference typically performs approximate Bayesian inference by maintaining a variational
posterior $q(s_{1:T})$ and minimising variational free energy
\begin{equation}
F[q]
= \mathbb{E}_{q(s_{1:T})}\big[\ln q(s_{1:T}) - \ln p(o_{1:T}, s_{1:T}\mid m)\big].
\end{equation}
\noindent
Minimising $F[q]$ tightens a bound on surprise $-\ln p(o_{1:T}\mid m)$ and makes
$q(s_{1:T})$ approximate $p(s_{1:T}\mid o_{1:T}, m)$.

\paragraph{Equivalent decomposition (energy--entropy form).}
\begin{equation}
F[q]
= \underbrace{\mathbb{E}_{q}\big[-\ln p(o_{1:T}, s_{1:T}\mid m)\big]}_{\text{expected energy}}
\;-\;
\underbrace{\mathbb{E}_{q}\big[-\ln q(s_{1:T})\big]}_{\text{entropy of } q}.
\end{equation}

\paragraph{Posterior update (perception).}
\begin{equation}
q^*(s_{1:T})
= \arg\min_{q} F[q].
\end{equation}
\noindent
In practice this is implemented via gradient flows, Laplace/VL updates, variational message passing,
or other approximate inference schemes.

\paragraph{Policies and expected free energy (planning).}
Let $\pi$ denote a policy (a sequence of future actions).
Active Inference selects policies by minimising expected free energy
\begin{equation}
G(\pi)
= \mathbb{E}_{q(o_{\tau}, s_{\tau}\mid \pi)}\big[\ln q(s_{\tau}\mid \pi) - \ln p(o_{\tau}, s_{\tau}\mid m)\big],
\end{equation}
\noindent
where $\tau$ indexes future time points and $q(o_{\tau}, s_{\tau}\mid \pi)$ is the policy-conditioned predictive density.

\paragraph{Common decomposition of expected free energy.}
A widely used decomposition is
\begin{equation}
G(\pi)
=
\underbrace{\mathbb{E}_{q(o_{\tau}\mid \pi)}\big[-\ln p(o_{\tau})\big]}_{\text{risk (preference violation)}}
\;+\;
\underbrace{\mathbb{E}_{q(o_{\tau}\mid \pi)}\big[H\!\left(p(o_{\tau}\mid s_{\tau})\right)\big]}_{\text{ambiguity}}
\;-\;
\underbrace{\mathbb{E}_{q(o_{\tau}\mid \pi)}\big[ \mathrm{IG}(s_{\tau};o_{\tau}\mid \pi)\big]}_{\text{epistemic value}},
\end{equation}
\noindent
where $p(o_{\tau})$ encodes prior preferences over outcomes, $H(\cdot)$ is entropy, and $\mathrm{IG}$ denotes information gain
(e.g., $\mathrm{IG}(s;o)=\mathrm{KL}(q(s\mid o)\|q(s))$). Exact forms vary with factorisation assumptions.

\paragraph{Policy posterior with precision.}
Policies are typically selected using a softmax (Boltzmann) distribution
\begin{equation}
p(\pi)
= \sigma\!\left(-\beta\, G(\pi)\right)
\;\;\propto\;\;
\exp\!\left(-\beta\, G(\pi)\right),
\end{equation}
\noindent
where $\beta$ is an inverse temperature (policy precision) controlling stochasticity of policy selection.

\subsection*{Polyphonic Active Inference (multiple generative models / ``voices'')}

\paragraph{Ensemble of generative models (voices).}
Assume a set of $K$ generative models (voices)
\begin{equation}
\mathcal{M} = \{m_1, m_2, \dots, m_K\}.
\end{equation}
\noindent
Each voice $k$ supports its own latent-state posterior $q_k(s_{1:T})$ under its own assumptions
(e.g., precisions, priors, dynamics, observation mappings).

\paragraph{Voice-specific variational free energy.}
\begin{equation}
F_k[q_k]
= \mathbb{E}_{q_k(s_{1:T})}\big[\ln q_k(s_{1:T}) - \ln p(o_{1:T}, s_{1:T}\mid m_k)\big].
\end{equation}
\noindent
Each voice can be updated using the same inference machinery as conventional Active Inference
(e.g., Laplace/VL, message passing), but applied independently within each $m_k$.

\paragraph{Non-dominating integration: polyphonic free energy.}
Polyphonic intelligence combines local objectives while penalising destructive inconsistency:
\begin{equation}
\mathcal{F}_{\mathrm{poly}}
=
\sum_{k=1}^{K} \pi_k \, F_k[q_k]
\;+\;
\sum_{i<j} \lambda_{ij}\, C\!\left(q_i, q_j\right).
\end{equation}
\noindent
Here, $\pi_k \ge 0$ are \emph{credence weights} (with $\sum_k \pi_k=1$),
$C(q_i,q_j)$ is a consistency cost (soft alignment), and $\lambda_{ij}\ge 0$ are coupling strengths.
Crucially, there is no hard model selection (no winner-takes-all pruning).

\paragraph{Examples of consistency costs.}
A simple choice is alignment in predicted outcomes:
\begin{equation}
C\!\left(q_i, q_j\right)
=
\mathbb{E}_{q_i(o_{\tau})}\!\left[\phi(o_{\tau})\right]
-
\mathbb{E}_{q_j(o_{\tau})}\!\left[\phi(o_{\tau})\right]
\;\;\;\Rightarrow\;\;\;
C_{ij} = \left\|\mu_i - \mu_j\right\|^2,
\end{equation}
\noindent
where $\phi(\cdot)$ is a feature map (e.g., goal-relevant summaries) and $\mu_k$ denotes the corresponding predicted feature mean.
Alternative choices include $\mathrm{KL}$ divergences between predictive distributions, or penalties on incompatible latent factors.

\paragraph{Polyphonic inference (local updates under coupling).}
A generic coupled update can be written as
\begin{equation}
q_k^*
=
\arg\min_{q_k}
\left[
\pi_k\, F_k[q_k]
+
\sum_{j\neq k} \lambda_{kj}\, C\!\left(q_k, q_j\right)
\right],
\end{equation}
\noindent
which reduces to standard Active Inference when $K=1$ and all coupling terms vanish.

\paragraph{Voice-specific expected free energy (planning).}
Each voice evaluates policies using its own predictive density:
\begin{equation}
G_k(\pi)
=
\mathbb{E}_{q_k(o_{\tau}, s_{\tau}\mid \pi)}\big[\ln q_k(s_{\tau}\mid \pi) - \ln p(o_{\tau}, s_{\tau}\mid m_k)\big].
\end{equation}
\noindent
As in the single-model case, $G_k(\pi)$ can be decomposed into risk, ambiguity, and epistemic value
with respect to the voice’s generative assumptions.

\paragraph{Polyphonic policy value (non-dominating integration for control).}
Define a population-level policy objective
\begin{equation}
G_{\mathrm{poly}}(\pi)
=
\sum_{k=1}^{K} \pi^{\mathrm{ctrl}}_k
\Big(
G_k(\pi)
+
\lambda\, C_k(\pi)
\Big),
\end{equation}
\noindent
where $\pi^{\mathrm{ctrl}}_k$ are \emph{control influence weights} (not necessarily equal to $\pi_k$),
and $C_k(\pi)$ is a policy-level alignment penalty. For example, aligning on a goal-progress statistic $r_k(\pi)$:
\begin{equation}
C_k(\pi)
=
\left(r_k(\pi) - \bar{r}(\pi)\right)^2,
\qquad
\bar{r}(\pi)=\sum_{j=1}^{K} \pi_j\, r_j(\pi).
\end{equation}
\noindent
This encourages agreement on coarse progress signals while allowing persistent disagreement elsewhere.

\paragraph{Decoupling credence from control.}
A simple convex mixing that prevents executive domination is
\begin{equation}
\pi^{\mathrm{ctrl}}_k
=
\alpha\, \pi_k
+
(1-\alpha)\,\frac{1}{K},
\qquad 0\le \alpha \le 1,
\end{equation}
\noindent
so that even minority voices retain some influence over action selection.

\paragraph{Policy posterior under polyphonic EFE.}
\begin{equation}
p(\pi)
\propto
\exp\!\left(-\beta_{\mathrm{eff}}\, G_{\mathrm{poly}}(\pi)\right).
\end{equation}
\noindent
This retains the standard Active Inference softmax form, but replaces $G$ with $G_{\mathrm{poly}}$.

\paragraph{Precision as diplomacy (adaptive action precision).}
In polyphonic settings, commitment can be controlled by precision as a function of cross-voice agreement:
\begin{equation}
\beta_{\mathrm{eff}}(t)
=
\beta_0 \, f\!\left(\mathcal{A}(t)\right),
\end{equation}
\noindent
where $\mathcal{A}(t)$ is an agreement (or coalition) index and $f(\cdot)$ is monotone increasing.
One simple choice uses the variance of predicted progress across voices:
\begin{equation}
\mathcal{A}(t)
=
-\frac{1}{|\Pi|}\sum_{\pi\in\Pi}\mathrm{Var}_k\!\big(r_k(\pi,t)\big),
\qquad
\beta_{\mathrm{eff}}(t)
=
\mathrm{clip}\!\left(\beta_0 \exp\!\big(\kappa\,\mathcal{A}(t)\big), \beta_{\min}, \beta_{\max}\right),
\end{equation}
\noindent
where $\Pi$ is the candidate policy set and $\mathrm{clip}$ bounds precision.

\paragraph{Updating credence weights from model evidence (without collapse).}
Instead of hard model selection, credence can be updated via a soft evidence accumulator:
\begin{equation}
\ell_k(t)
=
\rho\,\ell_k(t-1) - F_k(t),
\qquad
\pi_k(t)
=
\epsilon \frac{1}{K}
+
(1-\epsilon)\,
\frac{\exp(\gamma\,\ell_k(t))}{\sum_{j=1}^K \exp(\gamma\,\ell_j(t))},
\end{equation}
\noindent
where $\ell_k(t)$ is a leaky log-evidence proxy, $\rho\in(0,1)$ sets the timescale,
$\gamma$ controls sharpness, and $\epsilon$ enforces a floor (pluralism guarantee).

\paragraph{Viability.}
Let $\mathcal{V}$ denote a viability set over internal and external states (e.g., physical constraints, bounded energy,
bounded uncertainty, bounded coupling). Polyphonic control can be cast as maintaining viability by modulating coupling and precision:
\begin{equation}
\mathcal{V}=\{x:\; g_r(x)\le 0 \ \forall r\},
\qquad
\dot{\lambda}_{ij} = h_{ij}\!\left(\text{slack}(x)\right),
\qquad
\dot{\beta}_{\mathrm{eff}} = u\!\left(\text{slack}(x), \mathcal{A}(t)\right),
\end{equation}
\noindent
so that the system becomes more decisive (higher coupling/precision) near constraint boundaries
and more plural/exploratory (lower coupling/precision) when safely within $\mathcal{V}$.

\paragraph{Reduction to conventional Active Inference.}
The polyphonic formulation reduces to standard Active Inference under any of the following conditions:
\begin{itemize}
\item $K=1$ (a single generative model),
\item $\lambda_{ij}=0$ for all $i,j$ (no coupling),
\item $\pi_k = \pi^{\mathrm{ctrl}}_k = \delta_{k,k^*}$ (hard model selection),
\item or $\beta_{\mathrm{eff}}=\beta_0$ is fixed and independent of inter-voice agreement.
\end{itemize}
In these limits, $\mathcal{F}_{\mathrm{poly}} \rightarrow F$ and $G_{\mathrm{poly}}(\pi) \rightarrow G(\pi)$, recovering the standard Active Inference scheme.\\

Interpretational summary.
Polyphonic Active Inference preserves the normative objective of free energy minimisation, while relaxing the organisational assumption that inference and control must be governed by a single dominant generative model. Multiple models remain concurrently viable, coordination is achieved through soft alignment rather than elimination, and commitment is regulated by adaptive precision. In this sense, polyphony specifies a mode of inference--control organisation compatible with Active Inference, rather than an alternative objective function.

\end{document}